\documentstyle[11pt,epsf]{article}
\begin{document}
\baselineskip 0.2in
\title{Noise in Genotype Selection Model\footnote{The project supported by National Natural
Science Foundation of China (Grant No. of 10275099) and GuangDong
Provincial Natural Science Foundation (Grant No. of 021707 and
001182) }}
\author{Bao-Quan  AI$^{1}$, Wei  CHEN$^{2}$ , Xian-Ju
WANG$^{1}$,  Guo-Tao  LIU$^{1}$,\\De-Hua Wen$^{1,3}$and Liang-Gang
LIU$^{1}$} \maketitle
\begin{center}
$^{1}$Department of Physics, ZhongShan University,
GuangZhou, P.R.China\\
$^{2}$Department of Physics, JiNan  University, GuangZhou,
P.R.China.\\
$^{3}$Department of Physics, South China University of
technology,\\GuangZhou, P. R. China.\\
\end{center}
\parskip 0pt

\begin{abstract}
\baselineskip 0.25in We study the steady state properties of a
genotype selection model in presence of correlated Gaussian white
noise. The effect of the noise on the genotype selection model is
discussed. It is found that correlated noise can break the balance
of gene selection and induce the phase transition
which can makes us select one type gene haploid from a gene group.
Pacs: 87. 10. +e, 05. 40. -a, 02. 50. Ey.
\end{abstract}
\vskip 0.2in
 \baselineskip 0.25in
\section {Introduction}
\parskip 0pt
\indent Recently, nonlinear stochastic systems with noise terms
have attracted extensive investigations.  The concept of
noise-induced transition has got wide applications in the field of
physics, chemistry and biology \cite{1}\cite{2}\cite{3}. In most
of these theories the noise affects the dynamics through system
variable, i.e., the noise is multiplication in nature\cite{4}. The
focal theme of these investigations is steady state properties of
systems in which fluctuations, generally applied from outside, are
considered independent of the system's characteristic dissipation.
These studies consider only one type of fluctuations that affect
system's parameters, i.e., they drive multiplicatively. However,
fluctuations due to some extrinsic factors affect the system
directly , i.e., they drive system dynamics additively. Since two
types of fluctuations have the common origin, they correlated with
each other in the relevant timescale of the problem\cite{5}. On
the level of a Langevin-type description of a dynamical system,
the presence of correlation between noise can change the dynamics
of the system\cite{6}\cite{7}.  Correlated noise processes have
found applications in a broad range of studies such as steady
state properties of a single mode laser \cite{8}, bistable
kinetics\cite{9}, directed motion in spatially symmetric periodic
potentials\cite{10}, stochastic resonance in linear
systems\cite{11}, and steady state entropy production\cite{12}. In
this paper we study the genotype selection model in presences of
the correlated noise and discuss how noise correlation can break
the balance of the gene selection and induce the phase
transition.\\
\section {The simple genotype selection  model}
\indent We select a haploid group as our object and suppose that
each haploid may have gene $A$ or gene $B$\cite{13}. The number of
gene $A$ haploid,
 gene $B$ haploid and the total are $N_{A}$,$N_{B}$ and
 $N$,respectively. Because the total number is constant
 ($N=N_{A}+N_{B}$=constant), we can make a transform as follow:\\
\begin{equation}\label{e1}
 x=\frac{N_{A}}{N}, 1-x=\frac{N_{B}}{N}, 0\leq x\leq1.
\end{equation}
\indent $x$ is the ratio of gene $A$ number to the total number.
If gene $A$ haploid and gene $B$ haploid may have a mutation
($A\longrightarrow B$ or $B \longrightarrow A$) during the process
of the heredity and the ratio of $A\longrightarrow B$ and $B
\longrightarrow A$ are $m_{A}\Delta t $ and $m_{B}\Delta t$,
respectively, we can get the following difference
equation:\\
\begin{equation}\label{e2}
x(t-\Delta t)-x(t)=-m_{A}\Delta t x(t)+m_{B}\Delta t [1-x(t)]
\end{equation}
\indent On the other hand, on account of self-sow each gene
haploid has its rebirth rate:
\begin{equation}\label{e3}
N_{A}(t+\Delta t)=\omega_{A}N_{A}(t), N_{B}(t+\Delta
t)=\omega_{B}N_{B}(t)
\end{equation}
Where $\omega_{A}=1+\frac{S_{t}}{2},
\omega_{B}=1-\frac{S_{t}}{2}$, $S_{t}$ is selection gene and
$\Delta t$ is time gap between border generations. From Eq.(1) and
Eq.(3) we can get:
\begin{equation}\label{e4}
 x(t-\Delta t)-x(t)=\frac{S_{t}x(t)[1-x(t)]
 }{1-\frac{S_{t}}{2}+S_{t}x(t)}.
\end{equation}
\indent Considering Eq.(2) and Eq.(4) together we can get the
differential equation of $x(t)$ at the case of $\Delta
t\longrightarrow 0$.
\begin{equation}\label{e5}
\dot{x}=\beta-\gamma x+\mu x(1-x).
\end{equation}
\indent Where $\beta=m_{B}, \gamma=m_{A}+m_{B},
\mu=\frac{S_{t}}{\Delta t}$. In order to simplify the equation, we
suppose $m_{A}+m_{B}=1$. So the simplified gene selection dynamic
equation is shown as following \cite{14}
\begin{equation}\label{e6}
 \dot{x}=\beta-x+\mu x(1-x).
\end{equation}
\indent Now if due to some  environmental external disturbance the
gene selection rate of the haploid, it is likely to affect both
$\beta$ and $\mu$ in the form of additive and multiplicatively
noises that are connected through a correlation parameter. In
other words the external fluctuations affect the parameter $\mu$
which fluctuates around a mean value, thus generating
multiplicative noise and at the same time environmental
fluctuations perturbs the dynamics directly which gives
rise to additive noise. So We have\\
\begin{equation}\label{e7}
\dot{x}=\beta-x+\mu x(1-x)+x(1-x)\epsilon (t)+\Gamma (t).
\end{equation}
\indent Where $\epsilon (t), \Gamma (t)$ are Gaussian white noises
with the following properties.
 \begin{equation}\label{e8}
<\epsilon(t)>=<\Gamma(t)>=0,
\end{equation}

\begin{equation}\label{e9}
<\epsilon(t)\epsilon(t^{'})>=2D{\delta}(t-t^{'}),
\end{equation}

 \begin{equation}\label{e10}
<\Gamma(t)\Gamma(t^{'})>=2\alpha{\delta}(t-t^{'}),
\end{equation}

 \begin{equation}\label{e11}
 <\epsilon(t)\Gamma(t^{'})>=2\lambda\sqrt{D\alpha}{\delta}(t-t^{'}),
 \end{equation}
\indent where $D,\alpha$ are the strength of noise $\epsilon(t)$
and $\Gamma(t)$, respectively.
 $\lambda$ denotes the degree of correlation between noises $\epsilon(t)$
 and $\Gamma(t)$ with $0\leq\lambda<1$.\\

\section {Steady State Analysis of the Model}
\indent We can derive the corresponding Fokker-Planck equation for
evolution of Steady  Probability Distribution Function (SPDF)based
on Eq.(7)-Eq.(11). The equation is shown as follow. \cite{15}.\\
 \begin{equation}\label{e12}
 \frac{\partial P(x,t) }{\partial t}=-\frac{\partial A(x)P(x,t)}{\partial x}
 +\frac{\partial^{2}B(x)P(x,t)}{\partial x^{2}}.
\end{equation}
Where
\begin{equation}\label{e13}
A(x)=2Dx^{3}-(\mu+3D)x^{2}+(D+\mu-1-2\lambda\sqrt{D\alpha})x+\beta+\lambda
\sqrt{D\alpha}.
\end{equation}
\begin{equation}\label{e14}
B(x)=D[x(1-x)]^{2}+2\lambda\sqrt{D\alpha}x(1-x)+\alpha.
\end{equation}
\indent The steady probability distribution of Fokker-Planck
equation is given by \cite{14}
\begin{equation}\label{e15}
 P_{st}(x)={N_{0}\over B(x)}\exp[\int^{x}\frac{A(x^{'})}{B(x^{'})}dx^{'}]
\end{equation}
\indent Where $N_{0}$ is the normalization constant, using the
forms of $A(x)$ and $B(x)$ we have the  following integral forms
of the SPDF of Eq.(12)\cite{16}.
\begin{equation}\label{e16}
P_{st}(x)=\frac{N_{0}}{\sqrt{D[x-x^{2})]^{2}+2\lambda\sqrt{D\alpha}(x-x^{2})+\alpha}}
\exp[\int^{x}\frac{[\beta-x^{'}+\mu (x^{'}-{x^{'}}^{2})]dx^{'}
}{D(x^{'}-{x^{'}}^{2})^{2}+2\lambda\sqrt{D\alpha}(x^{'}-{x^{'}}^{2})+\alpha}]
\end{equation}
\indent Now we will give a numerical analysis for Eq. (16) and the
results will be presented as Fig. 1-Fig. 4.

\begin{figure}[htb]
\centerline{\epsfxsize 10cm \epsffile{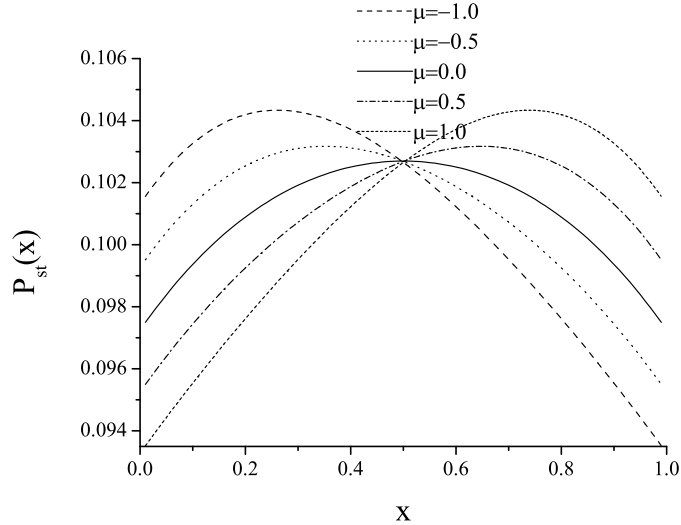}}\label{fig1}
\caption{Plot of $P_{st}(x)$ (denotes the probability) against x
(the ratio of gene $A$ number to the total) for different values
of the selection gene $\mu$.
$D$=0.5,$\alpha$=2.0,$\lambda$=0.5,$\beta$=0.5 and
$\mu$=-1.0,-0.5,0.0,0.5 and 1.0, respectively. (units are
relative)}
\end{figure}

\indent In the Fig.1, we study the effect of the selection rate
$\mu$ on SPDF. When the selection rate is zero ($\mu=0$), the
curve is
 symmetry. It is said that gene $A$ haploid ($x$) and gene $B$
 haploid ($1-x$) have the same probability distribution. If the
 selection rate get a negative value the peak of the curve is
 biased to the left. It is evident that the selection is unfair
 which is propitious to gene $B$ haploid. On the other hand, the
 selection is propitious to gene $A$ in case of $\mu >0$. In order
 to study the effect of the noise on the system easily we adopt
 the symmetrical case for the following discussion.\\

\begin{figure}[htb]
\centerline{\epsfxsize 10cm \epsffile{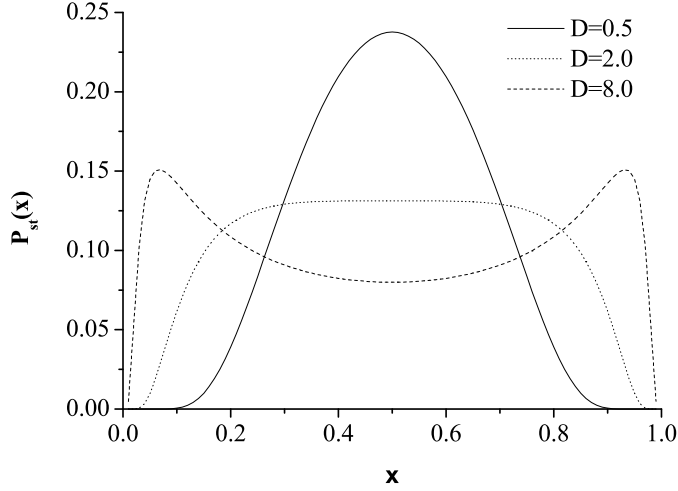}}\label{fig2}
\caption{Plot of $P_{st}(x)$ (denotes the probability) against x
(the ratio of gene $A$ number to the total) for different values
of multiplicative noise intensity $D$.
$\mu$=0,$\alpha$=0,$\lambda$=0,$\beta$=0.5 and $D$=0.5,2.0 and
8.0, respectively. (units are  relative)}
\end{figure}

  \indent In Fig.(2) we show the effect of multiplicative noise on
  SPDF. For a small value of $D$ the curve shows a single peak (at
  $x=0.5$) region which indicates that environmental selection
  gives the same chance to both gene $A$ haploid and gene $B$
  haploid and it is not easy for us to select one type haploid
  from the group. As the value of $D$ increases, the single peak
  region vanishes and it evolves a double peak region. The left
  peak is near the position $x=0$ while the right one is near the
  position $x=1$, which shows that environment selection gives a
  absolute big chance to one of the two (gene $A$ or gene $B$).
  In this case it is easy to select one type haploid from the
  group since the other type haploid number can be neglected
   with regard to the selected type number. On the other hand,
   from the Fig.2 we can know that the multiplicative noise
   can induce the phase transition from one  peak to double
   peak.\\
  \begin{figure}[htb]
\centerline{\epsfxsize 10cm \epsffile{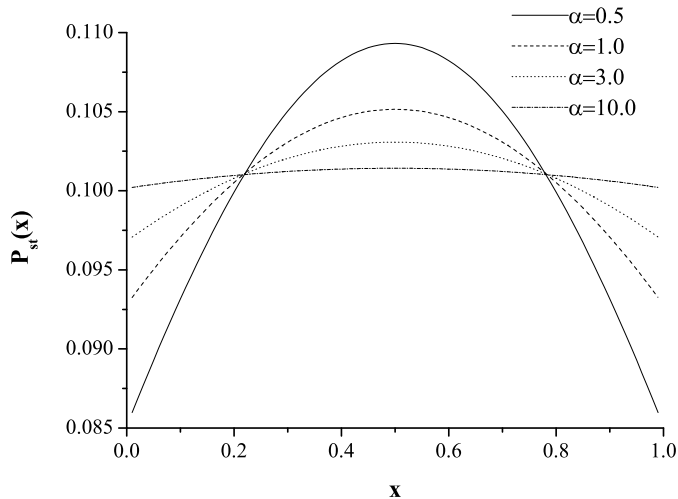}}\label{fig3}
\caption{$P_{st}(x)$ (denotes the probability) against x (the
ratio of gene $A$ number to the total) for different values of
additive noise intensity $\alpha$.
$\mu$=0,$D$=0,$\lambda$=0,$\beta$=0.5 and $\alpha$=0.5,1.0,3.0 and
10.0, respectively. (units are  relative)}
\end{figure}
 \indent The effect of the additive noise on SPDF is shown as
Fig.3. For a small value of $\alpha$ the curve represents a single
peak region which  vanishes for larger value of $\alpha$. The
  position of the peak is weakly affected by the strength of
  $\alpha$, however, its height may be affected  intensively by
  $\alpha$. It is said that the additive noise can not separate
  the single peak while it can make the peak disappear as a
  diffuse term.\\
  \begin{figure}[htb]
\centerline{\epsfxsize 10cm \epsffile{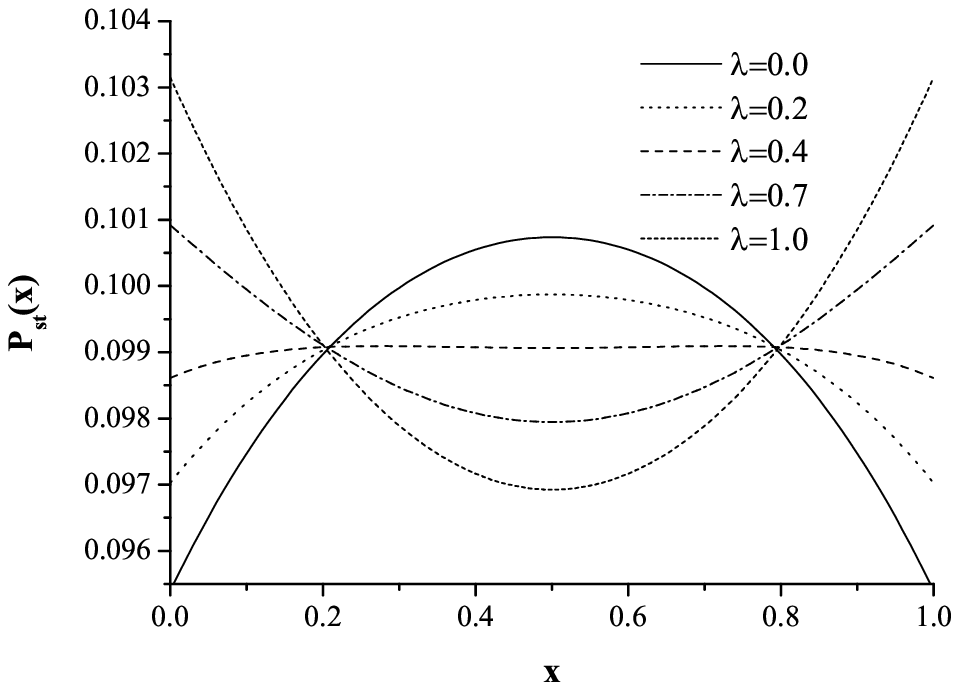}}\label{fig4}
\caption{$P_{st}(x)$ (denotes the probability) against x (the
ratio of gene $A$ number to the total) for different values of
noise correlation parameter $\lambda$.
 $\mu$=0.0,$\alpha$=2.0,$D$=0.5,$\beta$=0.5 and
$\lambda$=0.0,0.2,0.4,0.7 and 1.0, respectively. (units are
relative)}
\end{figure}

  \indent The curve of the Fig.4 shows the effect of correlation parameter
  on SPDF.  When $\lambda$ is zero, namely no correlation, the
  curve gives one peak region whose peak position is at $x=0.5$
  which shows that the selection is fair to both gene $A$ haploid
  and gene $B$ haploid. As the value of $\lambda$ increases the
  peak at $x=0.5$  vanishes  gradually and finally it becomes a
  concave  peak. At the same time, the height of the curve at the
  both side becomes larger gradually. In others word, the noise
  correlation make the probability  distribution centralize on
  both sides ($x=0$ and $x=1$). It is evident that the noise
  correlation make it  possible to separate one type haploid from
  a haploid group.\\
\section {Conclusion and Summary}
\indent In this paper, the steady state properties of the gene
selection model is investigated in presence of the correlated
Gaussian white noise. We study the effect of the white noise on
SPDF in the symmetrical case. The multiplicative noise can break
the single peak state to the double peak state (see Fig.2), which
indicates that the selection in the case is preponderant for one
of the two ( gene $A$ or gene $B$). The additive noise play a
diffusing role in the process, which causes the probabilities
distribution to a equal distribution. The noise correlation can
also break the symmetry of
$A$$B$, which induces the especial gene haploid to be selected.\\
  \indent From the above, it is found that the noise can change
  the nature selection from a equal probability selection to a
  differential probability one, which benefits us to select one type
  haploid from the haploid group. The breaking of the symmetry and
  the especial gene selection are very important to produce of gene order
  and biology evolution.\\
  \indent On the other hand, the stochastic force can induce the
  phase transition in our system. This viewpoint is completely novel
  as the traditional viewpoint thinks the stochastic force
  disturbs the phase transition.\\

\newpage

\end{document}